# A Voxel-Wise Uncertainty-Guided Framework for Glioma Segmentation Using Spherical Projection-Based U-Net and Localized Refinement in Multi-Parametric MRI

**Short Running Title: Uncertainty-Guided Glioma Segmentation**

Zhenyu Yang[1,2 †], Chen Yang[1,2 †], Rihui Zhang[1,2], Minbin Chen[3], Chunhao Wang[4], Fang-Fang Yin[1,2]

1. Medical Physics Graduate Program, Duke Kunshan University, Kunshan, Jiangsu, China
2. Jiangsu Provincial University Key (Construction) Laboratory for Smart Diagnosis and Treatment of Lung Cancer, Kunshan, Jiangsu, China
3. Department of Radiotherapy and Oncology, The First People's Hospital of Kunshan, Kunshan, Jiangsu, China
4. Department of Radiation Oncology, Duke University, Durham, NC, United States

[†] Equally Contributed
\* Corresponding Authors:
Zhenyu Yang
Medical Physics Graduate Program
Duke Kunshan University, Kunshan, Jiangsu, China
E-mail: zy84@duke.edu

# Abstract

*Purpose:* Accurate segmentation of glioma subregions from multi-parametric MRI (MP-MRI) is critical for diagnosis, treatment planning, and outcome monitoring, yet remains challenging due to tumor heterogeneity, poor inter-slice continuity, and ambiguous tissue boundaries. This study proposes an uncertainty-guided hybrid segmentation framework that integrates spherical projection-based 2D modeling and region-specific 3D refinement to improve segmentation fidelity and interpretability.

*Methods:* The proposed model was demonstrated using the BraTS2020 dataset, including a total of 369 patients with a four-modality MP-MRI protocol: FLAIR, T1, T1 contrast-enhanced (T1ce), and T2. For each patient, three classic 2D U-Nets were developed to segment the enhancing tumor (ET), tumor core (TC), and whole tumor (WT) respectively. The voxel-wise uncertainty map (i.e., quantitively measure the model confidence in segmentation results) was subsequently quantified using the 2D nnU-Net with spherical projection deformation, which characterizes the inconsistencies in segmentation predictions across multiple deformed input images, and image regions where the segmentation accuracy is limited in the initial 2D model can be highlighted. A 3D sliding window kernel was then implemented to capture the imaging volume with high uncertainty. Based on the collected kernels, a 3D nnU-Net model was trained to provide focused segmentation refinements to the high-uncertainty imaging volumes. The 3D segmentation results were then integrated with the 2D segmentation using the weighted average method. Particle Swarm Optimization was employed to optimize the weights of 2D and 3D segmentation results. The standalone 2D and 3D nnU-Net models were also evaluated as comparison studies. Voxel-wise accuracy, sensitivity, specificity, and 3D dice similarity coefficient were adopted to evaluate the segmentation performance.

*Results:* The proposed method significantly outperformed standalone 2D and 3D baselines across ET, TC, and WT targets. Notably, it achieved a Dice score of 0.8124 for ET (vs. 0.7527 for 2D and 0.6530 for 3D), 0.7499 for TC (vs. 0.7002 for 2D and 0.7027 for 3D), 0.9055 for WT (vs. 0.8989 for 2D and 0.9038 for 3D) and demonstrated consistent gains in sensitivity and accuracy. Visualizations confirmed improved spatial coherence, boundary preservation, and robustness in structurally complex regions.

*Conclusion:* This study introduces a novel uncertainty-aware segmentation framework that adaptively combines 2D and 3D predictions guided by interpretable uncertainty maps. By localizing computational effort to regions of ambiguity, the proposed approach enhances segmentation accuracy and efficiency, offering a promising solution for precision neuro-oncology and generalizable to other high-stakes medical imaging tasks.

# 1. Introduction

Gliomas represent one of the most aggressive and prevalent forms of primary malignant brain tumors in adults, originating from glial cells within the central nervous system[1,2]. Glioblastoma multiforme (GBM) is one of its aggressive subtypes with heterogeneous morphology, highly invasive growth patterns, and significant resistance to standard therapeutic interventions[1,2]. In clinic, accurate delineation and segmentation of glioma subregions are critically important for precise diagnosis, effective therapeutic planning, and reliable longitudinal monitoring. For example, reliable segmentation enables neurosurgeons to optimize the extent of tumor resection, assists radiation oncologists in accurately defining radiotherapy target volumes, and provides clinicians with objective metrics for assessing treatment response and disease progression[3–5]. Currently, multi-parametric magnetic resonance imaging (MP-MRI) has been widely accepted as the clinical standard imaging modality for glioma diagnosis and evaluation. Due to its superior soft-tissue contrast, MP-MRI effectively characterizes tumor morphology, cellular composition, and peritumoral edema. Nevertheless, accurate segmentation of gliomas from MP-MRI data remains challenging[3–5]. Traditional manual segmentation methods, despite their widespread clinical use, are labor-intensive, prone to substantial inter- and intra-observer variability, and difficult for routine or large-scale clinical deployment[6–8].

Recent advances in artificial intelligence have led to the widespread development of deep learning (DL)-based approaches for automated glioma segmentation. Among them, convolutional neural networks (CNNs), particularly U-Net and its variants, have achieved strong performance on benchmark datasets such as BraTS by leveraging multi-modal MP-MRI inputs[6–8]. Most of these models, including recent large-scale generalist frameworks such as the Segment Anything Model (SAM), perform segmentation in a 2D slice-by-slice manner[9,10]. The final volumetric segmentation is typically reconstructed by stacking predictions across slices. This design simplifies model training and enables deployment with limited GPU memory, but it fails to capture spatial continuity along the axial (z-axis) direction. As a result, such models are prone to discontinuities between adjacent slices, misalignment at tumor boundaries, and anatomically implausible segmentations[11]. These limitations become particularly pronounced in glioma subregions with subtle contrast differences or irregular shapes[3,4,12–14]. For example, the non-enhancing tumor core often exhibits low signal intensity on both T1 and T1ce, making it difficult to delineate consistently in 2D. Similarly, peritumoral edema seen on FLAIR may span several slices with gradual intensity transitions, which are easily missed without 3D context[15,16]. Although 3D CNNs such as V-Net and 3D U-Net have been proposed to address these issues by directly modeling volumetric context[17–19], they introduce high computational burden and are sensitive to voxel anisotropy, particularly in clinical scans where slice thickness is coarser than in-plane resolution. In addition, the relatively small tumor-to-brain volume ratio further complicates 3D learning due to severe class imbalance and limited availability of high-quality 3D annotations. These practical constraints continue to limit the robustness, efficiency, and clinical applicability of existing DL-based glioma segmentation solutions.

To address the limitations of current segmentation methods, recent research has increasingly focused on uncertainty quantification as a strategy to improve both algorithmic reliability and human-machine interaction[12,13,20–24]. Unlike traditional segmentation models that produce deterministic binary masks, uncertainty-aware frameworks generate voxel-wise confidence scores, offering more informative outputs that highlight regions of potential error. These voxel-level uncertainty maps serve as indicators of prediction confidence and can guide clinicians to focus their review on ambiguous or high-risk areas, thus improving both efficiency and reliability in clinical workflows[22]. Technically, uncertainty in deep learning is broadly categorized into two types: epistemic and aleatoric. Epistemic uncertainty reflects uncertainty in the model parameters due to insufficient or biased training data and can be estimated through Bayesian neural networks[25] or Monte Carlo (MC) dropout[26]. In MC dropout, multiple forward passes with randomly dropped neurons are used to simulate a posterior distribution over model predictions. Aleatoric uncertainty, in contrast, arises from inherent noise in the input data, such as poor tissue contrast or motion artifacts, and is often estimated using test-time augmentation (TTA)[27]. TTA introduces controlled perturbations to the input, such as flipping, rotation, or intensity scaling, and quantifies the variation in output predictions across augmented samples.

Building on these concepts, our study recently introduced a U-Net architecture with spherical projection-based image deformation to quantify segmentation uncertainty in brain tumor MRI[12,13]. Unlike methods that estimate uncertainty from feature distributions or latent activations, this technique operates directly in the input domain. Each 2D axial slice is nonlinearly projected onto a set of spherical surfaces to amplify local geometric features while preserving global structure. The projected spherical images are independently segmented by a shared 2D nnU-Net, and the variance across the resulting segmentations is used to derive uncertainty. Empirically, the produced uncertainty maps consistently highlight clinically relevant failure-prone regions, such as subtle tumor margins, necrotic cores with heterogeneous signal, and peritumoral edema[12,13]. Building on this foundation, the present work proposes a novel uncertainty-guided hybrid segmentation framework designed to enhance the volumetric delineation of glioma subregions in MP-MRI. Departing from conventional strategies that apply 3D segmentation uniformly across the entire imaging volume, our approach exploits the previously generated uncertainty maps to localize and prioritize anatomically ambiguous regions for targeted refinement. The framework comprises three sequential stages. First, the spherical-projection U-Net performs slice-wise segmentation while simultaneously producing voxel-level uncertainty maps that reveal spatial zones susceptible to segmentation failure due to low contrast, structural complexity, or inter-slice inconsistency. Second, a kernel-based region selection algorithm identifies 3D sub-volumes exhibiting elevated cumulative uncertainty, thereby enabling resource-efficient localization of ambiguous areas. Third, targeted volumetric refinement is conducted within these sub-volumes using patch-based 3D segmentation models, which incorporate local spatial context to enhance segmentation accuracy. Collectively, this adaptive, uncertainty-driven pipeline represents a methodological shift from static segmentation architectures toward dynamic, interpretable workflows that offer improved robustness and clinical utility in the delineation of glioma substructures.

## 2. Materials and Methods

### A. Patient Data

The study employed a total of 369 glioma patients from Brain Tumor Segmentation Challenge 2020 (BraTS 2020) dataset[28–30]. For each patient, 4 MR images were acquired as an MP-MRI protocol, including T1-weighted, T1-weighted contrast-enhanced (T1ce), T2-weighted, and fluid-attenuated inversion recovery (FLAIR). All MR images were preprocessed following the BraTS standard pipeline: skull-stripping, rigid registration to a common anatomical space, and isotropic resampling to a voxel resolution of 1×1×1 mm³. To ensure consistent spatial dimensions, all volumes were zero-padded to a matrix size of 256×256×160. The manual tumor segmentation ground truth includes three non-overlapping regions: Gadolinium (Gd)-enhancing tumor (ET), peritumoral edema (ED), and non-ET core (NCR/NET). Following the BraTS 2020 studies, another three overlapping segmentation targets were employed to evaluate the segmentation algorithms: ET, tumor core (TC, the combination of ET and NCR/NET), and whole tumor (WT, the combination of ET, NCR/NET, and ED)[28–30]. All segmentations were identified by at least two experienced radiologist or physicists. Figure 1 shows an example for the relationship between ET, TC, and WT.

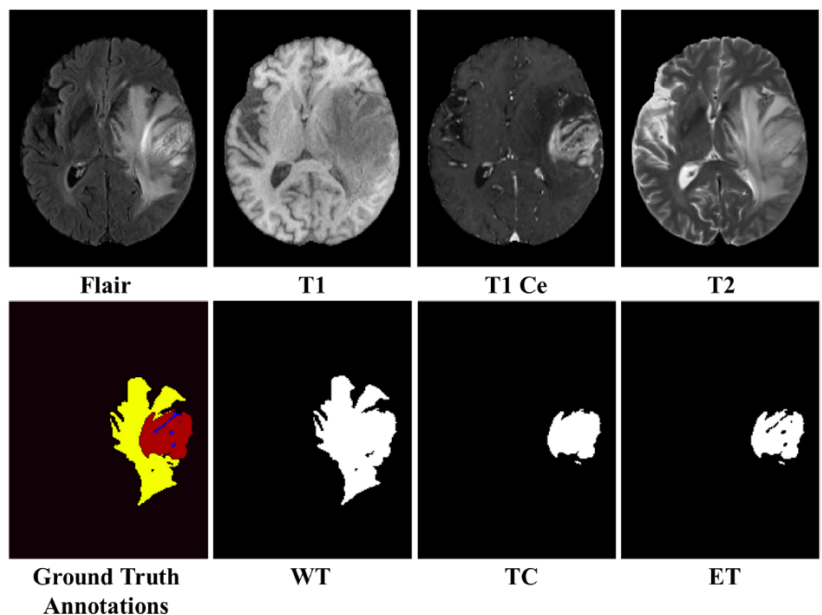

*Figure 1. An example of 4 MR images (i.e., FLAIR, T1, T1ce, T2) and the corresponding ground-truth segmentations (i.e., ET, TC, and WT) from the BraTS challenge 2020 dataset.*

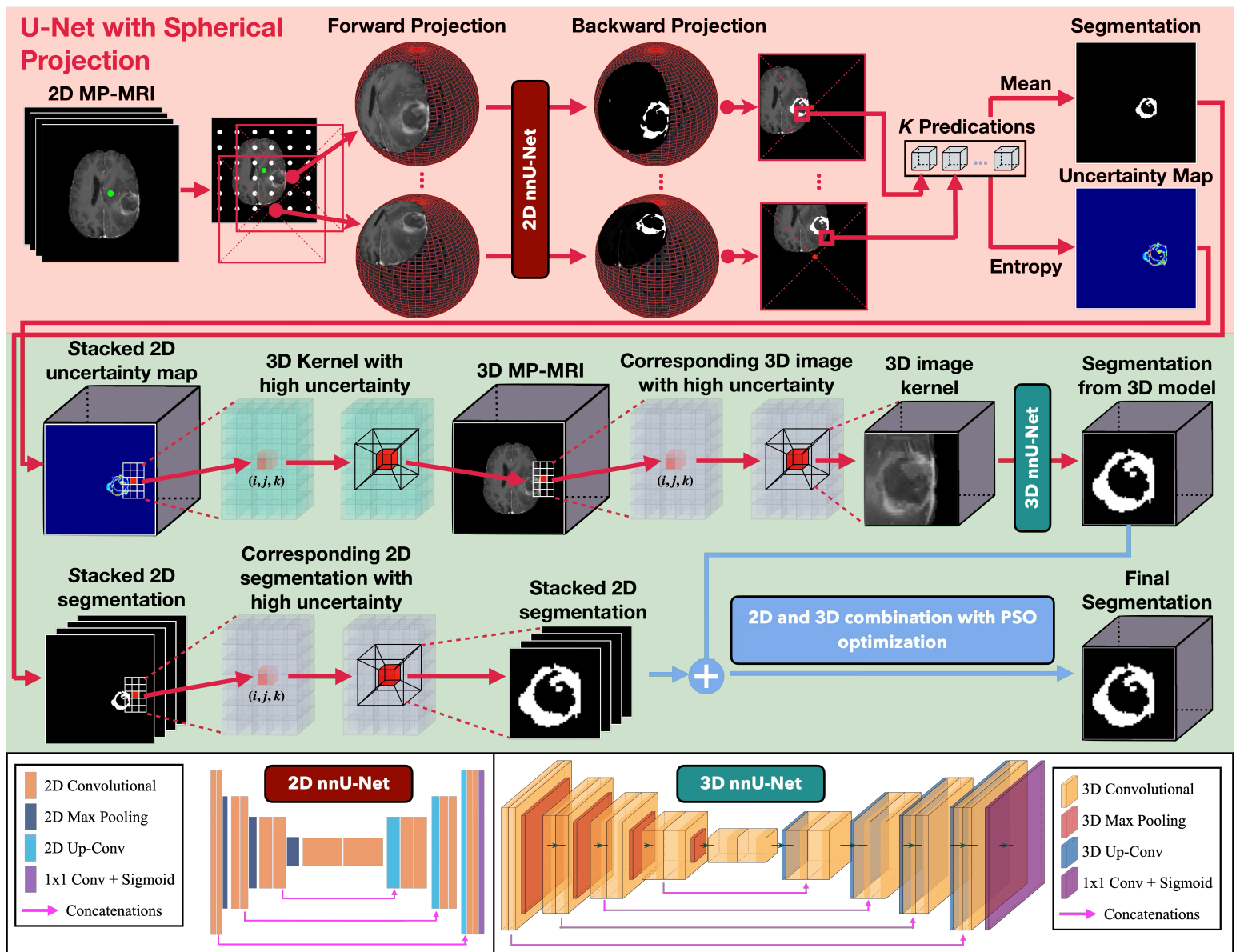

*Figure 2. Overview of the proposed uncertainty-guided hybrid segmentation framework.*

**B. Uncertainty-Guide Segmentation Model Design**

The architecture of the proposed uncertainty-guided segmentation framework is summarized in Figure 2. The model comprises three major components: (1) uncertainty quantification via spherical projection, (2) extraction of high-uncertainty subvolumes via kernel ranking, and (3) uncertainty-guided segmentation refinement through 3D deep learning and voxel-wise fusion. The framework is designed to integrate global context from 2D predictions and local precision from 3D volumetric refinement, while efficiently allocating computational resources to uncertain anatomical regions.

*B.1 Uncertainty Quantification via Spherical Projection*

The segmentation uncertainty quantifies the predictive confidence for each voxel's class assignment within the target segmentation region. Following our previous work[12,13], the voxel-level glioma segmentation uncertainty can be estimated using a geometric image transformation method inspired by spherical imaging processing[31]. As depicted in the top panel (red) of Figure 2, each 2D axial MR image slice was initially projected onto a series of virtual spherical surfaces, each associated with a unique projection origin. This spherical projection process introduces structured, non-uniform distortions across the image: anatomical details proximate to the projection origin are magnified, while the overall field-of-view and contextual integrity are largely preserved. To ensure comprehensive field-of-view coverage and transformation diversity, 1024 projection origins were uniformly distributed across the original 256×256-pixel MR image, with an 8-pixel interval in both dimensions, thereby generating 1024 distinct spherical projections for each 2D image slice.

Each spherically projected image was subsequently fed into a 2D nnU-Net architecture[32–35]. The nnU-Net is widely recognized as a state-of-the-art backbone for medical image segmentation due to its robust performance and adaptive capabilities. Its architecture is characterized by a symmetric encoder-decoder structure designed to capture both local spatial features and global contextual information. The encoder comprises multiple downsampling blocks, each consisting of two convolutional layers followed by instance normalization and Leaky ReLU activation, which effectively extract hierarchical features while progressively reducing spatial resolution.

The decoder mirrors this structure with a series of upsampling blocks, employing transposed convolutions and skip connections that bridge corresponding encoder layers to facilitate multi-scale feature integration. A final sigmoid activation layer outputs the probability (range [0,1]) of each pixel (within the spherical surface) belonging to a specific target class (e.g., ET, TC, or WT). A key architectural advantage of the nnU-Net, compared to conventional U-Net variants, is its automated adaptability to the specific characteristics of the input dataset[32,34]. Unlike manually designed networks, nnU-Net automatically configures architectural parameters such as input patch size, network depth, normalization type, and data augmentation strategies based on empirical data analysis. This self-configuring capability significantly enhances its robustness across diverse imaging modalities and anatomical targets.

For a given voxel $(i, j)$ in the original 2D MR image, 1024 independent segmentation predictions are obtained from the nnU-Net, corresponding to the 1024 spherical projected images. The initial 2D segmentation mask $P_{2D}(i, j)$ is derived by averaging these 1024 predictions, with a mean probability value exceeding 0.5 binarized as the target volume. Segmentation uncertainty $U_{2D}(i, j)$ was subsequently quantified by assessing the variability among these projected probability values using Shannon entropy[36]. Each continuous prediction value (range [0,1]) was first linearly discretized into 100 bins. The uncertainty $U_{2D}(i, j)$ is then calculated as:

$$U_{2D}(i,j) = -\sum_{t=1}^{T} f_{2D}^{(t)}(i,j) \cdot \ln\left(f_{2D}^{(t)}(i,j)\right) \quad (1)$$

where $f_{2D}^{(t)}(i,j)$ denote the empirical frequency of the $t$-th discretized bin across all projections for pixel $(i, j)$, and $T$ denotes the total number of unique non-zero bins in the pixel's uncertainty distribution. This entropy measure effectively captures prediction dispersion: low entropy indicates high consistency and confidence in model predictions, while high entropy reflects significant disagreement among projection-induced predictions, which is often localized near tumor boundaries or in structurally ambiguous regions. By processing the entire 3D MP-MRI volume slice-by-slice, this nnU-Net model with spherical projection deformation generates both the binarized segmentation mask $P_{2D}(i, j, k)$ and a corresponding voxel-wise uncertainty map $U_{2D}(i, j, k)$ for the entire volume. A more detailed description and implementation specification of the spherical projection segmentation model can be found in Reference[12,13].

*B.2 Kernel-Based Identification of High-Uncertainty Sub-volumes*

A kernel-based iterative search algorithm was subsequently developed to systematically identify 3D sub-volumes exhibiting high segmentation uncertainty. As illustrated in the middle panel (green) of Figure 2, a cubic sliding window kernel with a fixed dimension of $d \times d \times d$ mm³ was traversed across the entire uncertainty map $U_{2D}(i, j, k)$ in a dense grid. At each discrete position, the cumulative uncertainty within the enclosed region was computed as the sum of its constituent voxel-wise uncertainty values. The region yielding the highest cumulative uncertainty was initially selected as the primary candidate for further refinement. Subsequent candidate kernels were identified using the same procedure, with an additional constraint on permissible volumetric overlap between newly selected and previously chosen kernels. If a newly selected kernel exhibited a volumetric overlap with any previously chosen kernels exceeding this predefined constraint, it was discarded; otherwise, it was considered as a valid candidate for refinement. This iterative selection process continued until the cumulative uncertainty in all remaining unselected kernels approached a predefined minimum threshold, effectively identifying all significant high-uncertainty regions.

The kernel size ($d$) was treated as a critical hyperparameter influencing the granularity of uncertainty localization. A smaller kernel risks overfitting to localized noise or minor uncertainties, whereas an excessively large kernel may incorporate non-informative background, thereby compromising the precise localization of uncertainty. Based on a statistical analysis of tumor volumes within the BraTS 2020 dataset, the average volumes for ET, TC, and WT were estimated at approximately 10677 mm³ ($\approx 22^3$ mm³), 58389 mm³ ($\approx 39^3$ mm³), and 114361 mm³ ($\approx 49^3$ mm³), respectively. Accordingly, kernel sizes were empirically selected as $d = 32$ mm for ET and TC, and $d = 64$ mm for WT. These dimensions were chosen to align with typical anatomical extents of the respective tumor subregions, ensuring the inclusion of relevant structures within the high-uncertainty regions. Moreover, the volumetric overlap constraint is another crucial hyperparameter. As conceptually demonstrated in Figure 3, excessive overlap (e.g., 70%) between selected kernels results in redundant sub-volumes with highly similar anatomical content. Incorporating such highly overlapping kernels for subsequent refinement could potentially increase computational burden and lead to overfitting. Conversely, an insufficient overlap ratio (e.g., 10%) risks under-covering critical high-uncertainty regions, leading to incomplete refinement. In this study, we empirically set the volumetric overlap constraint to 40% of the entire 3D kernel volume ($d \times d \times d$ mm$^3$).

This iterative search algorithm effectively ranks potential kernel regions by cumulative uncertainty, prioritizing areas of higher uncertainty while ensuring diversity by rejecting redundant, highly overlapping candidates. For each patient, the selected 3D image kernels (for all four MRI modalities), along with their corresponding 3D ground truth kernels (for three respective tumor subregions), were extracted for subsequent refinement.

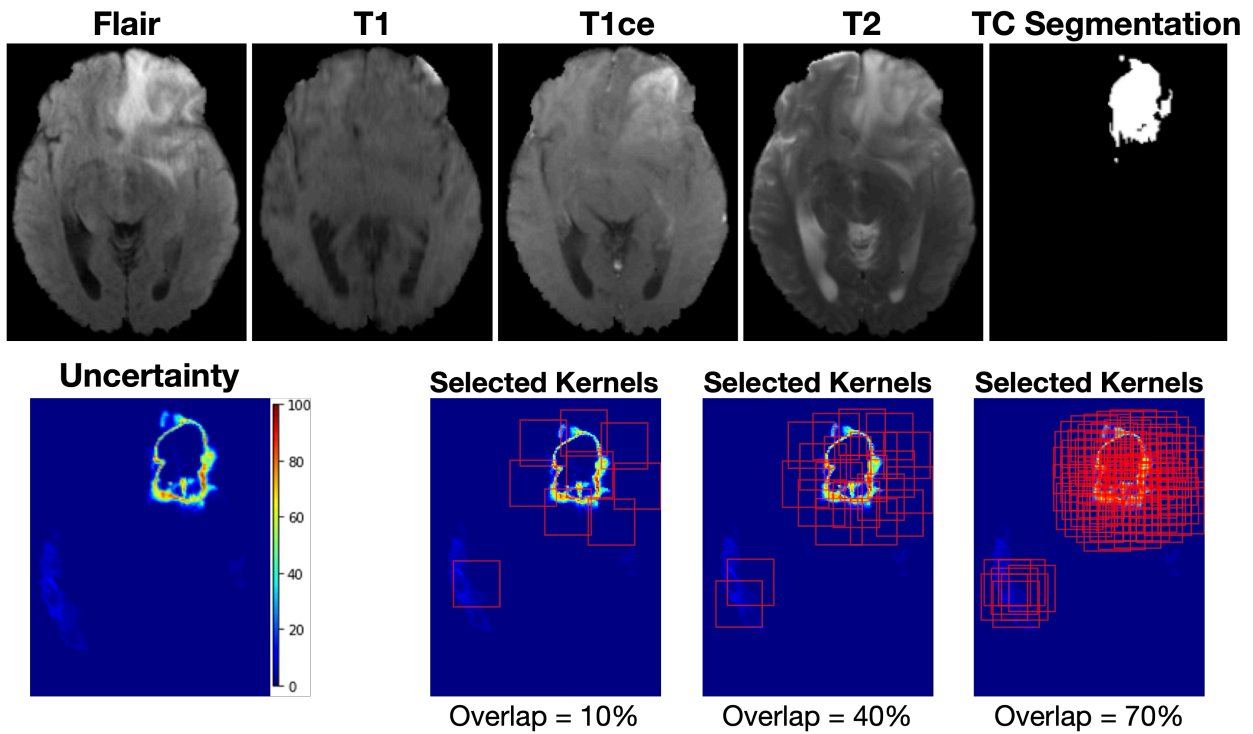

*Figure 3. Illustration of uncertainty-guided kernel selection with varying overlap ratios. Top row shows representative axial MP-MRI slices (FLAIR, T1, T1ce, T2) and ground-truth TC segmentation. The bottom row displays the voxel-wise uncertainty map generated by U-Net with spherical projection deformation and selected refinement kernels (red boxes) overlaid on the uncertainty map using three different overlap settings: 10%, 40%, and 70%.*

*B.3 Uncertainty-Guided Segmentation Refinement*

Each extracted 3D image kernel, identified as a high-uncertainty sub-volume, was subsequently fed into a dedicated 3D nnU-Net for localized segmentation refinement. This 3D nnU-Net shares an identical structural design with the 2D nnU-Net described previously, with the direct replacement of 2D convolutional layers, 2D pooling layers, and 2D upsampling layers with their 3D counterparts. This configuration enables the direct acquisition of 3D segmentation predictions with dimensions of $d \times d \times d$ mm³. This targeted training on selectively sampled uncertain regions allows the 3D model to concentrate its learning capacity on anatomically and clinically ambiguous areas, thereby improving local delineation. Consequently, for a given voxel $(i, j, k)$ in the original MR image, one or more refined 3D segmentation probability maps, denoted as $P_{3D_{local}}(i, j, k)$, are obtained from the 3D nnU-Net (allowing for overlapping predictions from distinct kernels).

In parallel, the original 2D segmentation prediction from the U-Net with spherical projection deformation, $P_{2D}(i, j, k)$, is also available for the same spatial location. These two predictions are then integrated using a voxel-wise fusion strategy designed to optimally combine the global contextual understanding from the 2D nnU-Net with the precise local refinement from the 3D nnU-Net. This fusion is governed by a sigmoid-weighted linear combination for each spatial location within the MP-MRI volume, expressed as:

$$P_{\text{fused}}(i, j, k) = \sigma\big(w_{2D} \cdot P_{2D}(i, j, k) + w_{3D} \cdot P_{3D\_composite}(i, j, k) + b\big) \quad (2)$$

Here, $p_{\text{fused}}$ is the final fused segmentation probability, $w_{2D}$ and $w_{3D}$ are weighting coefficients for the 2D and 3D models, respectively, $b$ is a bias term, and $\sigma$ denotes the sigmoid activation function, which transforms the combined linear output into a probability in the range [0,1] for binarized prediction. The fusion logic adapts dynamically based on 3D kernel coverage:

1. For voxels not covered by any extracted 3D high-uncertainty region, no 3D prediction is available. In these instances, the original 2D nnU-Net prediction is retained as the final segmentation output for that voxel.

2. For voxels covered by exactly one 3D kernel, $P_{3D\_composite}(i,j,k)$ is replaced by the single $P_{3D\_local}(i,j,k)$ prediction from that specific kernel, and fusion Equation (2) is applied.
3. In regions with multiple overlapping 3D kernels, multiple $P_{3D\_local}(i,j,k)$ predictions are available for a given voxel. These are averaged to form a composite 3D probability: $P_{3D\_composite}(i,j,k) = \frac{1}{N} \sum_{n=1}^{N} P_{3D\_local,n}(i,j,k)$, where $N$ is the number of overlapping kernels contributing predictions to voxel $(i,j,k)$. Subsequently, fusion Equation (2) is applied using this $P_{3D\_composite}$.

The fusion parameters $(w_{2D}, w_{3D})$ are critical for dynamically balancing the contributions of the 2D and 3D models. This ensures that the precise 3D contextual information from the refined sub-volumes is leveraged where local ambiguities are high, while relying on the computationally efficient 2D segmentation in less complex or non-uncertain regions. To determine their optimal values, the Particle Swarm Optimization (PSO) algorithm was employed[37,38]. In this framework, each particle represents a candidate solution defined by a vector $(w_{2D}, w_{3D})$, subject to the non-negativity constraint $w_{2D}, w_{3D} \geq 0$. The optimization objective was to maximize the Dice Similarity Coefficient (DSC) across the target tumor subregions (ET, TC, and WT) on the test set.

*B.4 Model Implementation and Evaluation*

The BraTS 2020 dataset was randomly partitioned into training and independent test sets using an 8:2 split ratio, resulting in 300 training cases and 69 independent test cases. The proposed uncertainty-guided segmentation models were implemented independently for each of the three segmentation targets: ET, TC, and WT. Initially, three 2D nnU-Net models with spherical projection were trained to generate 2D segmentation masks and associated uncertainty maps slice-wise across the multi-parametric MRI input volumes. Subsequently, high-uncertainty regions were extracted for each target separately using the kernel-based analysis described in Section B.2. Three separate 3D nnU-Net models were then trained using these extracted sub-volumes to segment ET, TC, and WT kernels. All components of the proposed framework—including model training, identification of high-uncertainty regions, development of 2D and 3D nnU-Net models, and PSO algorithm execution—were strictly confined to the training data to prevent data leakage. Training for both the 2D nnU-Net and 3D nnU-Net models was conducted using the PyTorch framework. A binary cross-entropy loss function was adopted, and optimization was performed using the Adam optimizer with an initial learning rate of $1 \times 10^{-5}$. A batch size of 10 was employed, and training sessions were guided by early stopping criteria based on the validation DSC to prevent overfitting.

Model performance was finally evaluated on the 69 independent test cases using four standard quantitative metrics: Dice Similarity Coefficient (DSC), voxel-wise sensitivity, specificity, and overall accuracy. These metrics were calculated separately for each target tumor subregion (ET, TC, and WT). All experiments were executed on a local workstation equipped with a 16-core Intel Core i7-13700KF CPU (3.4 GHz), 16 GB system RAM, and an NVIDIA GeForce RTX 4070 GPU with 12 GB VRAM.

## C. Comparison Study

To rigorously assess the effectiveness of the proposed uncertainty-guided segmentation framework, a series of comparative experiments were conducted against two benchmark models:

1) **2D nnU-Net with spherical projection (without 3D refinement):** This model evaluated the standalone performance of the spherical projection-based segmentation approach. It accepted the original 4-channel MP-MRI slice as input and applied spherical projection to generate 1024 (with 8×8 projection center intervals) deformed versions of each slice. Each projected image was independently processed to produce segmentation probability maps, which were subsequently averaged to yield the final segmentation output. Similarly, three 2D nnU-Net with spherical projection were trained independently for ET, TC, and WT, and predictions were stacked across slices to reconstruct the volumetric output.

2) **3D nnU-Net (Fully volumetric model)**: A 3D nnU-Net model served as a high-capacity volumetric benchmark. Three separate models, accepting the full 3D MP-MRI volume (with four modalities) as input, were trained to perform end-to-end volumetric segmentation for ET, TC, and WT, respectively.

For all comparative models, the training configurations were kept consistent with the proposed uncertainty-guided framework, including training/test set partitions, the use of binary cross-entropy loss, and the same optimizer and initial learning rate. The models were evaluated on the same independent test cases using DSC, voxel-wise sensitivity, specificity, and accuracy. Statistical comparisons between the proposed uncertainty-guided model and each comparison model were conducted using the Wilcoxon signed-rank test with a significance level of 0.05.

## 3. Results

Figure 4-6 present representative examples from the test set, demonstrating the effectiveness of the proposed uncertainty-guided refinement pipeline across multiple glioma subregions. For each case, the four MP-MRI modalities (FLAIR, T1, T1ce, and T2), manual ground truth, model-predicted uncertainty map, and segmentation results from the proposed fusion model, the 2D nnU-Net, and the 3D nnU-Net are displayed.

- In ET segmentation (Cases 1 and 2), the proposed model consistently outperformed both baselines, especially in challenging anatomical regions. In Case 1, the enhancing tumor is well visualized on T1ce and T2, but the 3D nnU-Net fails to capture its inferior boundary, and the 2D model, though more extensive, exhibits boundary irregularity. The uncertainty map identifies the region as ambiguous, guiding the fusion model to produce a delineation that better aligns with the ground truth. In Case 2, where the enhancing lesion is small and fragmented, both baseline models underperform—either by missing components or introducing discontinuity. The proposed model, informed by localized uncertainty, effectively recovers the full lesion extent through selective refinement.
- For TC segmentation (Cases 3 and 4), similar trends are observed. In Case 3, the tumor core exhibits heterogeneous intensity on T1 and T1ce, leading the 2D model to undersegment the core and the 3D model to oversegment peripheral necrotic regions. The uncertainty map highlights the central region as uncertain, prompting refinement. The proposed model produces a compact and accurate segmentation. In Case 4, the core is embedded within a large surrounding edema, which confuses the 2D model. The 3D model partially includes non-core tissue, while the proposed fusion approach leverages low uncertainty regions to isolate the core volume with improved specificity.
- WT segmentation (Cases 5 and 6) further demonstrates the robustness of the fusion model in delineating irregular and topologically complex tumor volumes. In Case 5, both 2D and 3D models yield incomplete or noisy segmentations, whereas the fusion model consolidates the lesion with improved spatial integrity. In Case 6, the tumor spans multiple lobes with heterogeneous morphology, resulting in both over- and under-segmentation from the baselines.

Quantitative results for all segmentation targets are summarized in Table I, where the "*" marks the statistically significant difference. The proposed model consistently achieved superior Dice coefficients, sensitivity, and overall segmentation performance across ET, TC, and WT, with negligible compromise in specificity or accuracy. For ET, the proposed method achieved a mean Dice coefficient of 0.8124 ± 0.3136, compared to 0.7527 ± 0.3551 for the 2D model and 0.6530 ± 0.3340 for the 3D model. Sensitivity was markedly improved (0.9297 ± 0.2108), while voxel-wise accuracy and specificity remained high (0.9999 ± 0.0002). TC segmentation yielded similar trends, with the fusion model achieving a Dice score of 0.7499 ± 0.2205, compared to 0.7002 and 0.7027 for the 2D and 3D baselines, respectively. Sensitivity also increased significantly to 0.7700 ± 0.2779, demonstrating improved detection of true tumor voxels. WT segmentation showed comparable performance across all models, with the fusion model attaining the highest Dice (0.9055 ± 0.0648), while accuracy and specificity remained consistent among methods. These results confirm that the proposed hybrid approach effectively integrates 2D precision and 3D contextual information to resolve ambiguities, particularly in the ET and TC regions where traditional models tend to struggle.

Following PSO, the optimal fusion weights assigned to the 2D and 3D components varied by subregion (see Table II): $w_{2D}$=0.5848 and $w_{3D}$=0.4152 for ET; $w_{2D}$=0.6980, $w_{3D}$=0.3020 for TC; and $w_{2D}$=0.3204, $w_{3D}$=0.6796 for WT. In terms of computational efficiency, the training time per model varied according to architectural complexity. Each target-specific 2D nnU-Net required approximately 24 hours to converge, while the corresponding 3D nnU-Net models demanded 48 to 72 hours of training time due to the increased complexity of volumetric patch-based learning and extensive data augmentation. At inference, the full pipeline—comprising spherical projection, 2D segmentation, uncertainty estimation, regional extraction, local 3D refinement, and fusion—was completed within 4 minutes per patient.

*Table I: Quantitative Segmentation Performance on BraTS 2020 Test Set (Mean ± Standard Deviation).*

|     |              | 2D Model        | 3D Model        | Ours             |
|-----|--------------|-----------------|-----------------|------------------|
| ET  | **3D Dice**      | 0.7527±0.3551   | 0.6530±0.3340   | 0.8124±0.3136*   |
|     | **Accuracy (%)** | 0.9999±0.0002   | 09991±0.0010    | 0.9999±0.0002*   |
|     | **Sensitivity**  | 0.9201±0.2163   | 0.5591±0.3741   | 0.9297±0.2108*   |
|     | **Specificity**  | 0.9999±0.0002   | 0.9997±0.0004   | 0.9999±0.0001*   |
| TC  | **3D Dice**      | 0.7002±0.3103   | 0.7027±0.2844   | 0.7499±0.2205*   |
|     | **Accuracy (%)** | 0.9949±0.0076   | 0.9957±0.0067   | 0.9952±0.0063    |
|     | **Sensitivity**  | 0.6481±0.3384   | 0.6914±0.3311   | 0.7700±0.2779*   |
|     | **Specificity**  | 0.9996±0.0005   | 0.9989±0.0017   | 0.9983±0.0022    |
| WT  | **3D Dice**      | 0.8989±0.0915   | 0.9038±0.0052   | 0.9055±0.0648    |
|     | **Accuracy (%)** | 0.9963±0.0032   | 0.9963±0.0025   | 0.9963±0.0030    |
|     | **Sensitivity**  | 0.8704±0.1353   | 0.9081±0.0857   | 0.9017±0.1082    |
|     | **Specificity**  | 0.9989±0.0013   | 0.9981±0.0017   | 0.9982±0.0019    |

*"*" Marks the statistically significant difference.*

*Table II: Optimized Fusion Weights for Each Tumor Subregion*

|     | $w_{2D}$ | $w_{3D}$ |
|-----|----------|----------|
| ET  | 0.5848   | 0.4152   |
| WT  | 0.3204   | 0.6796   |
| TC  | 0.6980   | 0.3020   |

**ET segmentation**

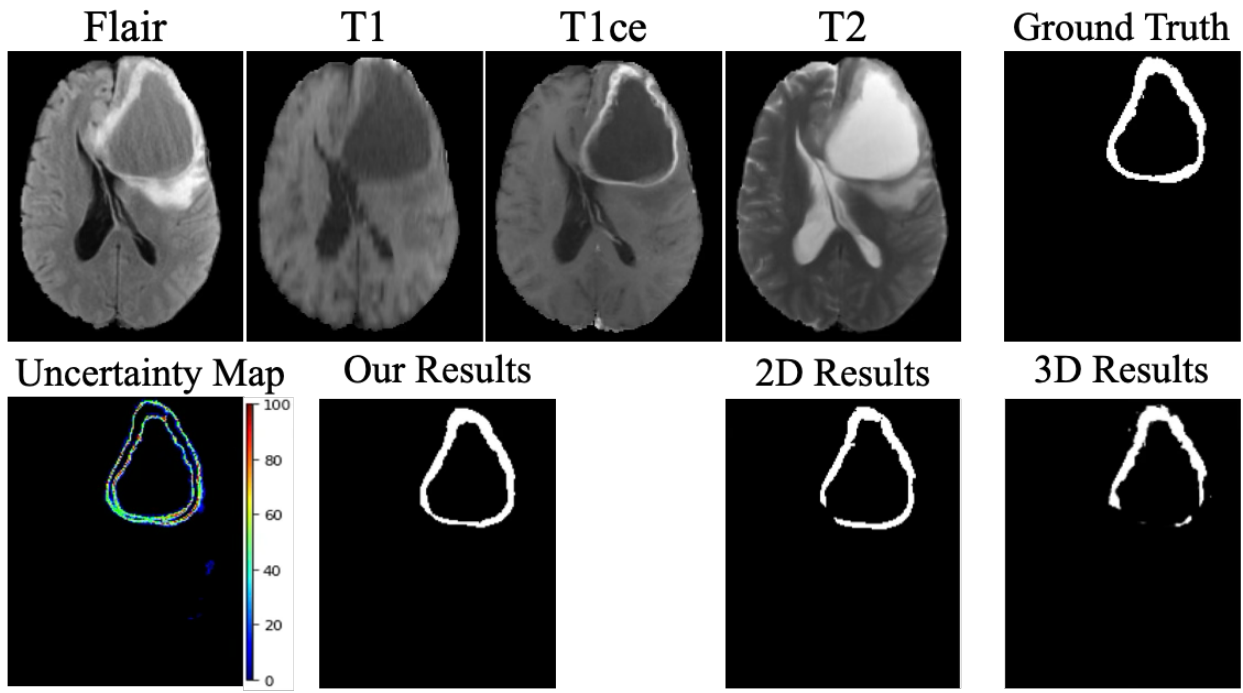

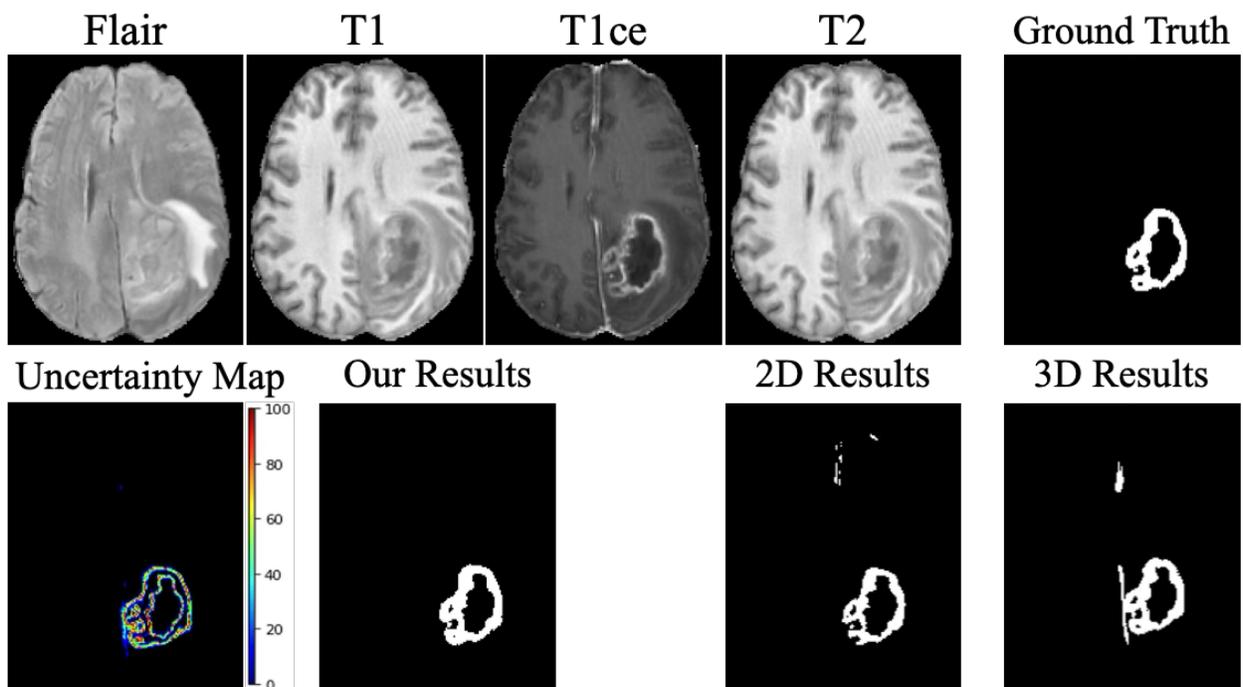

*Figure 4. Visual comparison of ET segmentation performance across representative cases.*

## TC segmentation

### Case 3

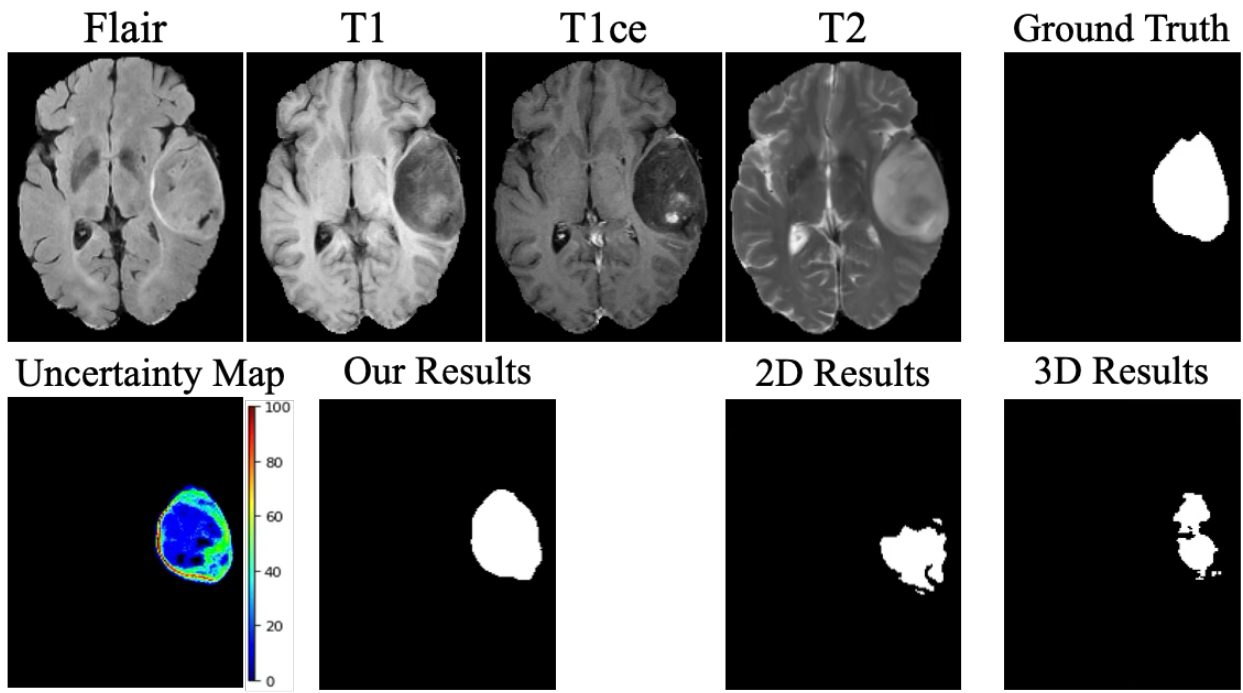

### Case 4

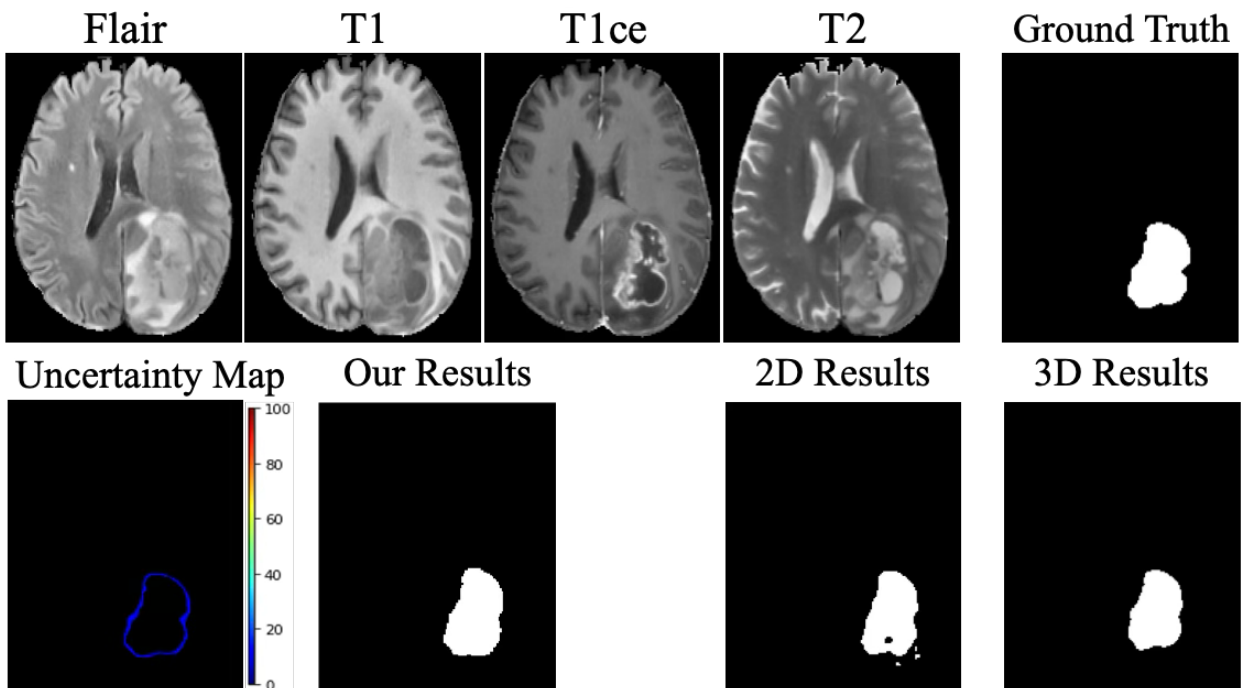

*Figure 5. Visual comparison of TC segmentation performance across representative cases.*

## WT segmentation

### Case 5

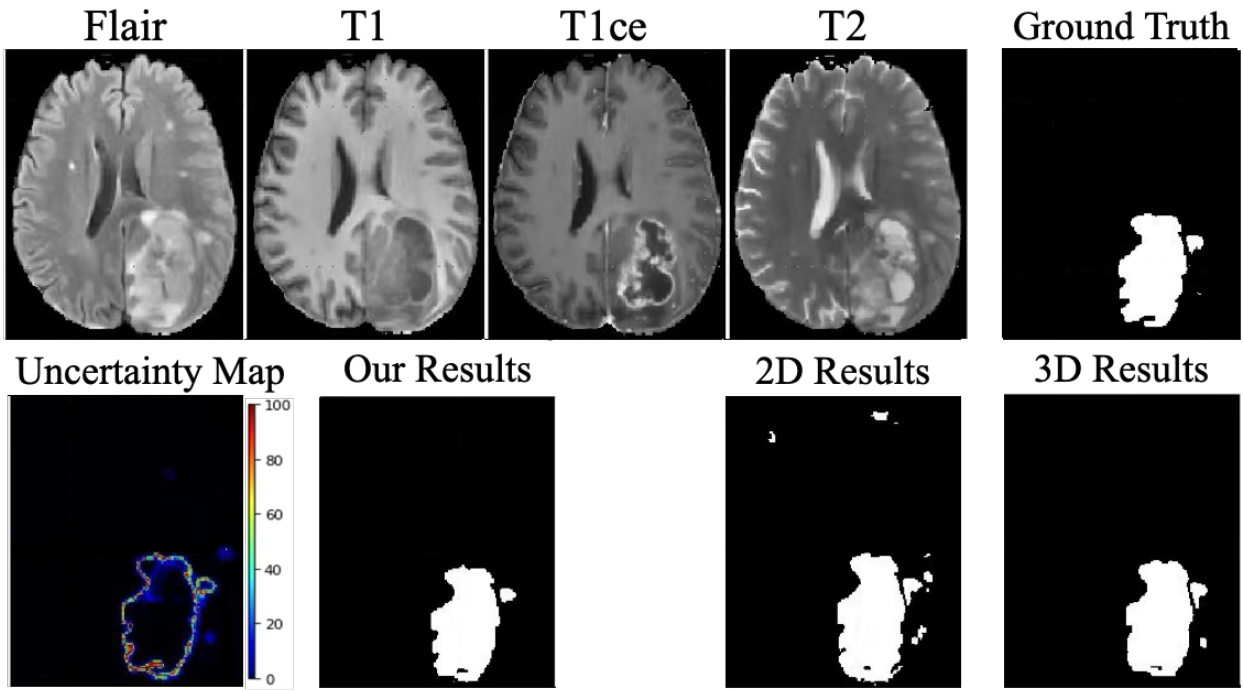

### Case 6

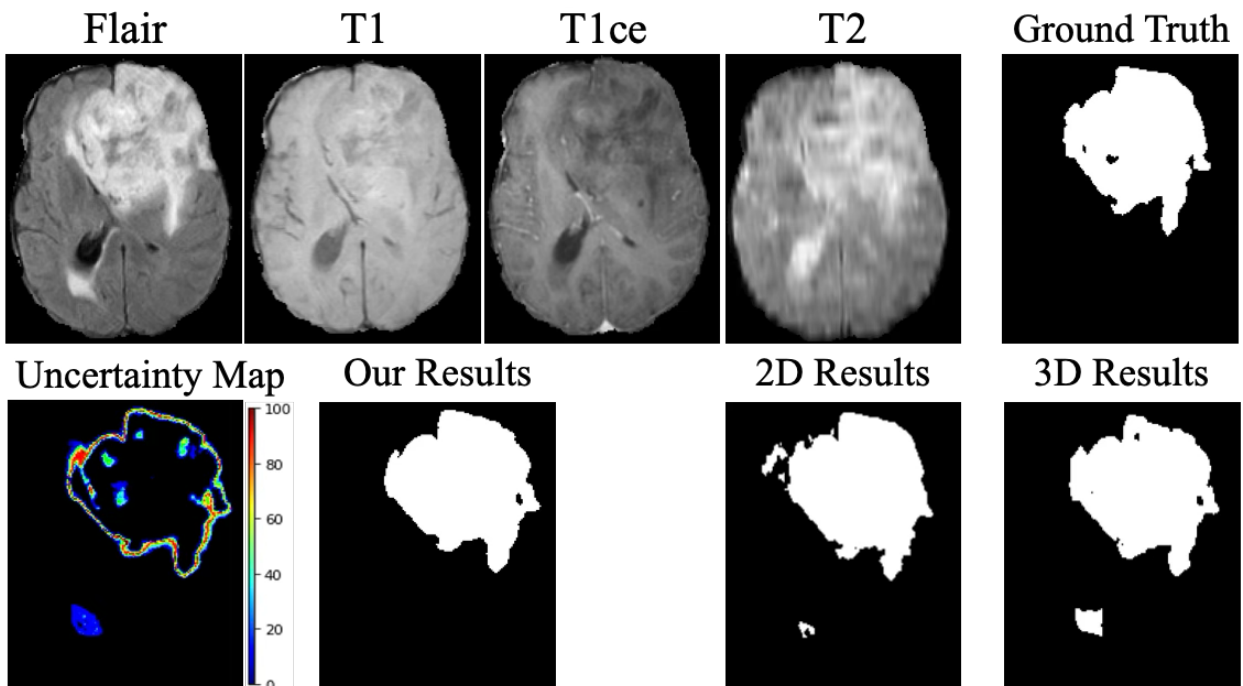

*Figure 6. Visual comparison of WT segmentation performance across representative cases.*

## 4. Discussion

In this study, we propose a hybrid segmentation framework that integrates spherical-projection-based uncertainty estimation with region-specific 3D refinement to enhance glioma segmentation in MP-MRI. While prior approaches have typically treated segmentation and uncertainty quantification as separate objectives, our framework explicitly unifies these tasks by leveraging voxel-wise uncertainty maps to identify regions of low model confidence and selectively applying targeted 3D refinement. The architecture consists of three interdependent stages: first, spherical projection-based 2D segmentation is employed to generate both preliminary tumor delineations and associated voxel-level uncertainty estimates; second, a ROI selection algorithm aggregates the uncertainty maps to identify spatially localized sub-volumes exhibiting high ambiguity; and third, a patch-based 3D segmentation model is applied within these regions to incorporate volumetric context and improve segmentation fidelity. This adaptive fusion strategy is specifically designed to mitigate two key limitations in existing glioma segmentation paradigms, i.e., (1) the inter-slice discontinuity common in 2D models and (2) the limited receptive field in patch-wise 3D networks. More broadly, the proposed framework aligns with an emerging consensus that uncertainty should not only serve as a post hoc evaluation criterion, but rather as an informative prior to guide model behavior, enabling more robust and interpretable solutions to complex segmentation tasks in clinical neuroimaging.

The previous proposed U-Net with spherical projection deformation forms the backbone of the initial segmentation and uncertainty quantification stage. By generating 1024 views per slice through nonlinear projections, the model encodes multiscale anatomical context without requiring architectural complexity. Prediction variance across these views provides a robust measure of voxel-wise uncertainty. As illustrated in Figure 4-6 (Cases 1-6), the generated uncertainty maps consistently highlight areas that are prone to segmentation failure, including tumor margins, necrotic centers, and low-contrast regions. For example, in Case 1 (ET), the uncertainty map reveals discontinuities along the tumor periphery that are missed by the 2D model. Similarly, in Case 3 (TC), heterogeneous intratumoral intensity leads to substantial disagreement among projections, aligning with visibly ambiguous regions on the ground truth. These examples confirm that the uncertainty signal is both anatomically and diagnostically meaningful, providing a reliable basis for downstream refinement.

Once uncertainty maps are computed, a kernel-based ranking algorithm is applied to identify sub-volumes requiring 3D refinement. This localized processing paradigm contrasts with full-volume 3D inference and focuses computational effort on diagnostically challenging regions. High-uncertainty ROIs are further processed using dedicated 3D nnU-Net models, and voxel-wise fusion between 2D and 3D outputs is performed within each sub-volume. Fusion weights are optimized using PSO. PSO can be a useful tool for the non-differentiable and non-convex objective functions like the Dice score, which can be difficult to be optimized using traditional gradient-based methods[37–41]. Given the fact that the low-dimensional and continuous nature of the weight space, PSO provides a simple yet powerful global optimization approach that avoids local minimum and does not require gradient information. As shown in Figure 6, this fusion strategy yields superior boundary continuity and region-wise consistency, particularly in regions where the 2D and 3D models independently fail. In Case 4 (TC), the 2D model omits the necrotic center, and the 3D model incorrectly includes peritumoral edema; their integration through PSO recovers a balanced, anatomically plausible prediction. This result illustrates that fusion is not merely an averaging step, but a critical operation governed by uncertainty-aware model complementarity. Our empirical evaluation (Table I) also confirms that this strategy leads to substantial improvements in segmentation accuracy, particularly in ET and TC subregions.

The region-specific fusion behavior observed in our framework is technically grounded in both anatomical characteristics and architectural capabilities, as reflected in the optimized PSO-derived weights (Table II). For enhancing tumor (ET) segmentation, the higher weight assigned to the 2D model ($w_{2D} = 0.5848 > w_{2D} = 0.4152$) aligns with the need for precise boundary localization in small, irregularly shaped, and contrast-enhanced regions typically captured in T1ce images. The 2D nnU-Net, leveraging spherical projections and high in-plane resolution, is particularly effective in detecting fine structures such as ring-enhancing margins or scattered nodular components, which are often smoothed over by the limited context and interpolation artifacts in patch-based 3D models. This preference is even more pronounced in tumor core (TC) segmentation ($w_{2D} = 0.6980 > w_{3D} = 0.3011$), where necrotic or non-enhancing cores exhibit high intensity contrast yet irregular morphology, making them more amenable to the multi-view sensitivity of the 2D model. In contrast, whole tumor (WT) segmentation, which includes extensive edema and infiltrative regions visible in FLAIR and T2 sequences, benefits more from the global context and interslice continuity of 3D models, as reflected by the dominant 3D weight ($w_{2D} = 0.3204 < w_{3D} = 0.6796$). The diffuse and low-amplitude uncertainty distributions in WT, illustrated in Figure 5 (Cases 5-6), suggest that the 2D predictions are already spatially consistent and confident, leaving little room for 3D correction to add value. This is further supported by the marginal Dice improvement in WT (0.9055 vs. 0.9038, Table I), in contrast to the significant gains observed in ET and TC. Together, these results demonstrate that the fusion strategy must be anatomically adaptive, with regional weighting reflecting the interplay between local structure complexity, contrast characteristics, and model-specific representational strengths.

Beyond improving segmentation accuracy, this study proposes a conceptually distinct approach that repositions uncertainty from a retrospective evaluation tool to an active component of the inference process. Rather than benchmarking against existing BraTS Challenge submissions, the focus here is on demonstrating a proof-of-concept framework in which uncertainty plays a functional role in guiding localized 3D refinement. By selectively invoking volumetric processing only in regions identified as uncertain through spherical projection-based analysis, the method avoids unnecessary computation in confidently segmented areas. This targeted design not only improves efficiency but also maintains segmentation quality across subregions with varying levels of complexity. As a result, the framework achieves an end-to-end inference time of under four minutes per subject, without requiring model simplification or resolution reduction. Such efficiency, combined with the modular and interpretable nature of uncertainty maps, makes the approach potentially adaptable for clinical scenarios with time constraints, including intraoperative settings and adaptive radiotherapy workflows.

Despite the demonstrated effectiveness, several technical limitations warrant further consideration. First, the current framework quantifies uncertainty using Shannon entropy derived from prediction variance across spherical projections. While this approach is computationally efficient and yields spatially meaningful maps, it may not fully capture both epistemic and aleatoric uncertainty[30]. More expressive modeling strategies, such as deep ensembles or Bayesian neural networks, could offer better calibration and robustness. Second, the region selection mechanism remains heuristically defined. Although the kernel-based uncertainty ranking algorithm effectively identified high-priority regions for refinement in this study, its reliance on empirically chosen parameters—such as kernel size *d*, overlap ratio, and stopping

thresholds—introduces tunable hyperparameters that may not transfer well across datasets or imaging conditions. Large kernels risk diluting uncertainty localization by including irrelevant voxels, whereas overly small kernels may fragment semantically coherent regions, impairing the contextual performance of the 3D model. A more adaptive strategy informed by uncertainty distribution statistics could yield a more principled and generalizable solution. Third, the framework was evaluated exclusively on the BraTS 2020 dataset. The generalizability of the proposed approach—particularly the utility of spherical projection for uncertainty estimation—remains untested in other clinical settings. Application to different tumor types, anatomical regions, or imaging modalities such as CT or PET may require projection scheme modification or fusion strategy adaptation.

## 5. Conclusion

In this study, we developed an uncertainty-guided segmentation framework that combines voxel-wise uncertainty estimation with targeted 3D refinement to enhance the accuracy and reliability of glioma segmentation from multi-parametric MRI data. By leveraging a 2D nnU-Net with spherical projection for initial segmentation and uncertainty quantification, and selectively applying 3D nnU-Net refinement to high-uncertainty regions identified via a kernel-based ranking algorithm, our method effectively allocates computational resources to anatomically ambiguous or low-confidence areas. This uncertainty-driven refinement strategy presents a computationally efficient, clinically relevant approach that can be generalized to other imaging modalities and anatomical structures.

## Conflict of Interest

None associated with this work

## Funding Statement

This work was supported by National Natural Science Foundation of China under Grant 12405382.